\documentclass[%
 journal=ancac3,
 manuscript=article,
]{achemso}

\usepackage[version=3]{mhchem} 

\usepackage{xr}
\makeatletter
\newcommand*{\addFileDependency}[1]{
   \typeout{(#1)}
   \@addtofilelist{#1}
   \IfFileExists{#1}{}{\typeout{No file #1.}}
}
\makeatother

\newcommand*{\myexternaldocument}[1]{
     \externaldocument{#1}
     \addFileDependency{#1.tex}
     \addFileDependency{#1.aux}
}
\myexternaldocument{SI}

\usepackage{gensymb}
\let\oldAA\AA
\renewcommand{\AA}{\text{\normalfont\oldAA}}

\usepackage[section]{placeins}
\makeatletter
\let\ftype@table\ftype@figure
\makeatother
\usepackage{chemmacros}
\usepackage{upgreek}

\setkeys{acs}{articletitle = true}

\title{Spin-Dependent Transport Through a Colloidal Quantum Dot: The Role of Exchange Interactions}

\author{John P. Philbin}
\affiliation{Harvard John A. Paulson School of Engineering and Applied Sciences, Harvard University, Cambridge, Massachusetts 02138, United States}
\email{jphilbin@g.harvard.edu}

\author{Amikam Levy}
\affiliation{Department of Chemistry, Bar-Ilan University, Ramat-Gan 52900, Israel}
\email{amikam.levy@biu.ac.il}

\author{Prineha Narang}
\affiliation{Harvard John A. Paulson School of Engineering and Applied Sciences, Harvard University, Cambridge, Massachusetts 02138, United States}
\email{prineha@seas.harvard.edu}

\author{Wenjie Dou}
\affiliation{School of Science, Westlake University, Hangzhou, Zhejiang 310024, China }
\alsoaffiliation{Institute of Natural Sciences, Westlake Institute for Advanced Study, Hangzhou, Zhejiang 310024, China}
\email{douwenjie@westlake.edu.cn}

\abbreviations{}


\begin{document}

\begin{tocentry}
\includegraphics[width=8.25cm]{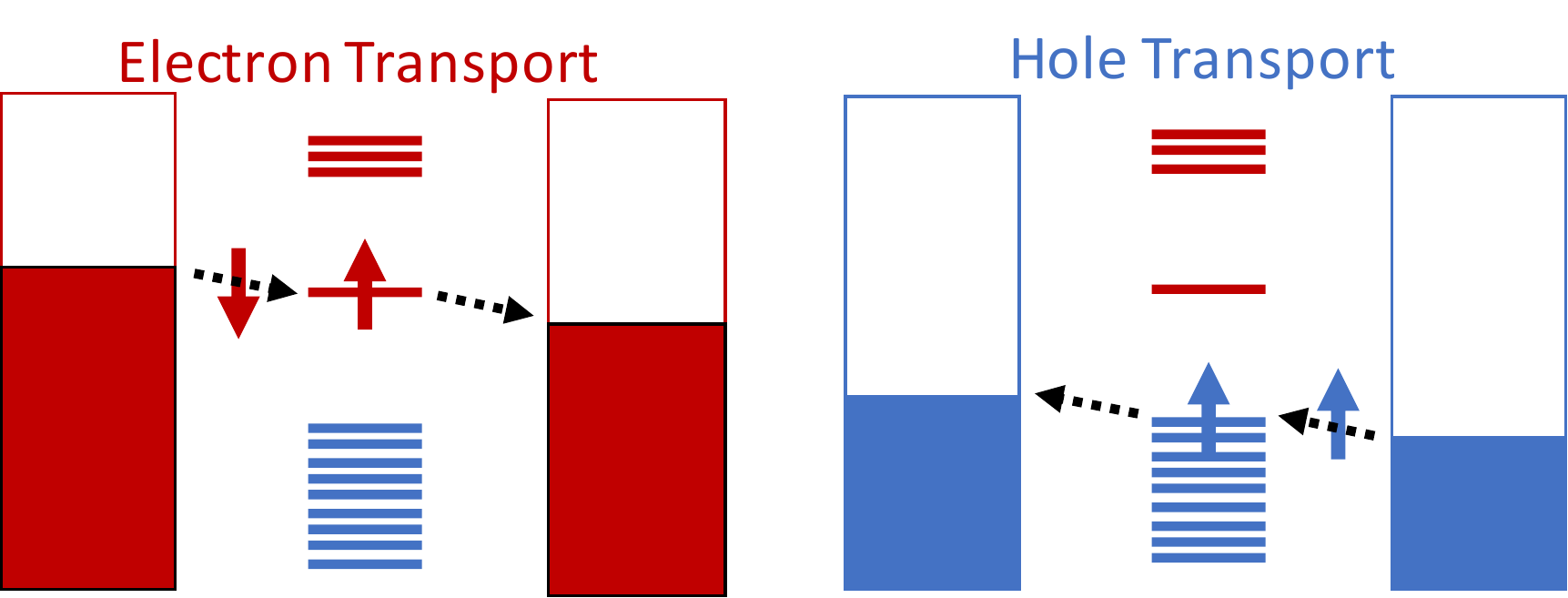}
\end{tocentry}

\begin{abstract}
The study of charge and spin transport through semiconductor quantum dots is experiencing a renaissance due to recent advances in nano-fabrication and the realization of quantum dots as candidates for quantum computing. In this work, we combine atomistic electronic structure calculations with quantum master equation methods to study the transport of electrons and holes through strongly confined quantum dots coupled to two leads with a voltage bias. We find that a competition between the energy spacing between the two lowest quasiparticle energy levels and the strength of the exchange interaction determines the spin states of the lowest two quasiparticle energy levels.  Specifically, the low density of electron states results in a spin singlet being the lowest energy two-electron state whereas, in contrast, the high density of states and significant exchange interaction results in a spin triplet being the lowest energy two-hole state. The exchange interaction is also responsible for spin blockades in transport properties, which could persist up to temperatures as high as $77$~K for strongly confined colloidal quantum dots from our calculations. Lastly, we relate these findings to the preparation and manipulation of singlet and triplet spin qubit states in quantum dots using voltage biases.
\end{abstract}

\subsection*{Introduction}

Semiconductor quantum dots (QDs) have been studied and utilized for a wide range of applications, including light emitting diodes,\cite{Colvin1994,Shirasaki2013} lasers,\cite{Pietryga2016} photocatalysis,\cite{Ben-Shahar2018} nano-junctions,\cite{Banin1999} and quantum information processing.\cite{Loss1998,Burkard1999,Imamoglu1999,Kloeffel2013,Veldhorst2015,West2019,Yang2020,Petit2020,Kagan2021} All of these applications require a detailed understanding of the excited states in these confined semiconductors. To this end, electron-hole pairs (i.e. excitons) have been thoroughly characterized via experimental and theoretical studies for both III-V and II-VI QDs in the weak and strong confinement regimes.\cite{Scholes2006,Sercel2018} Analogous to electron-hole pairs being of central importance to all applications involving the absorption and emission of light by QDs, two-electron (electron-electron) and two-hole (hole-hole) states are central to QD-based quantum computation, which now attracts great research interests.\cite{Loss1998,Burkard1999,Veldhorst2015,West2019,Yang2020,Petit2020} 

The use of two-electron or two-hole states as qubits requires the ability to initialize, manipulate, and detect the spin state of the two-quasiparticle states.\cite{Loss1998,Burkard1999} This has recently been done in a variety of QDs, primarily in weakly confined silicon and III-VI QDs.\cite{Petta2005,Veldhorst2015,Nichol2017,Yang2020,Petit2020,Chan2021} A crucial breakthrough that lead to these studies is that of rapid single-shot measurement techniques capable of distinguishing singlet and triplet states.\cite{Barthel2009,Dehollain2014,West2019} These measurements are key to the initialization and detection stages of QD-based quantum computation. To better design and understand these measurements, previous theoretical investigations utilized purely model Hamiltonians in transport studies through QDs. Specifically, the ``universal'' Hamiltonian of a QD in the large Thouless conductance limit is given by~\cite{Alhassid2003} 
\begin{eqnarray}
\hat H & = & \sum_{n\sigma}E_{n}\hat d_{n\sigma}^{\dagger}\hat d_{\lambda\sigma}+U\hat{n}^{2}-K_{S}\hat{\mathbf{S}}^{2},\label{eq:qd-model-hamiltonian}
\end{eqnarray}
where $E_{n}$ are spin-degenerate single-particle levels ($\sigma = \uparrow, \downarrow$ denotes spin-up or spin-down), $U$ is the Hubbard parameter representing the Coulomb repulsion, $\hat{n}$ is the particle-number operator, $\hat{\mathbf{S}}$ is the total-spin operator, and $K_{S}$ is a constant exchange interaction strength. While this model offers important physical intuition for understanding the transport properties,\cite{Baranger2000,Rosch2001,Eto2002,Alhassid2003} an atomistic electronic structure based investigation of these energies and their impact on charge and spin transport of electrons and holes through QDs remains elusive, until now. 

In this work, we combine atomistic electronic structure calculations with quantum master equation methods to study the low temperature charge and spin transport properties of strongly confined II-VI QDs. (We restrict ourselves to within the manifold of low QD occupation numbers in order to detail the intricate physics involved in the singlet-triplet two-electron and two-hole states from a microscopic point of view.) We find that the differences in the density of single-particle states ($\Delta_{\text{e,h}}\equiv E_{1}-E_{0}$) and many-body interactions ($U_{\text{e,h}}$ and $K_{\text{e,h}}$) lead to qualitative differences between the lowest energy two-electron and two-hole states. We show that  the strength of the exchange interaction and Hund's rules play an important role in the energetic ordering of singlet and triplet two-electron and two-hole states. We then demonstrate that the spin transport dynamics of electrons and holes through confined II-VI QDs are also different, highlighting the spin-dependent blockades that arise from the exchange interaction in semiconductor QDs. Lastly, we discuss the temperature dependence of these features along with their impact on the initialization of the singlet and triplet states that are central to QD-based quantum computation.

\begin{figure}[t]
\centering{}\includegraphics[width=12cm]{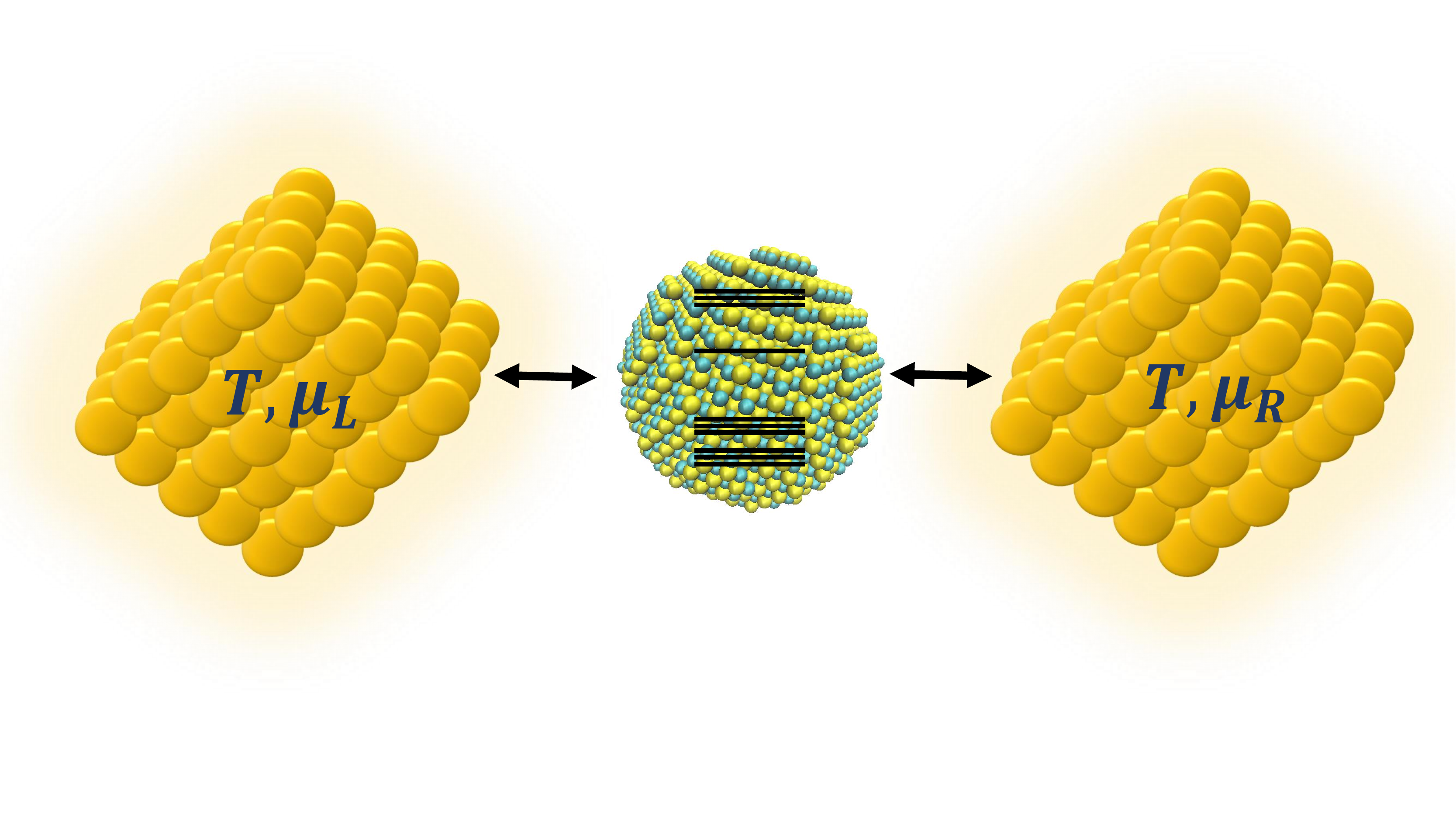} \caption{\label{fig:qd-transport-schematic}Schematic representation of left ($L)$ and right ($R$) leads (e.g. gold tips) being held at a constant common temperature ($T$) but different chemical potentials ($\mu_{L}$ and $\mu_{R}$, respectively). These leads are coupled to a single CdSe quantum dot (Cd and Se atoms are shown in yellow and green, respectively). The quantized single quasiparticle (i.e. electron and hole) energy levels of the CdSe quantum dot are represented by horizontal black lines.}
\end{figure}

\subsection*{Results and Discussion}

We now briefly introduce our computational methods. Details of the method are described in the Methods section. We first utilize the semi-empirical pseudopotential method to obtain the quasiparticle (i.e. electron and hole ) states.\cite{Wang1996,Rabani1999b} The Coulomb interactions between the quasiparticle states are given by
\begin{eqnarray}
W_{ijmn} & = & \int\int\,\psi_{i}^{*}\left(x\right)\psi_{j}^{*}\left(y\right)W\left(x,y\right)\psi_{m}\left(x\right)\psi_{n}\left(y\right)\,dx\,dy\,.\label{eq:coulomb-matrix-elements}
\end{eqnarray}
Here, $\psi_{n}$ is the wavefunction obtained by solving the semi-empirical pseudopotential Schr\"{o}dinger equation and $W$ is the screened Coulomb operator. Next, we build the Hamiltonian for the QDs (in second quantization notation):
\begin{eqnarray}
\hat{H}_{\text{QD}} & = & \sum_{n}E_{n}\hat{d}_{n}^{\dagger}\hat{d}_{n}+\sum_{ijmn}W_{ijmn}\hat{d}_{i}^{\dagger}\hat{d}_{j}\hat{d}_{n}^{\dagger}\hat{d}_{m}\label{eq:qd-hamiltonian}
\end{eqnarray}
where $\hat{d}_{n}$ ($\hat{d}_{n}^{\dagger}$) is the annihilation (creation) operator for quasiparticle state $\psi_{n}$ with energy $E_{n}$. 

In order to calculate charge and spin transport through the QD, we couple the QD to two leads, as shown schematically in \ref{fig:qd-transport-schematic}. We set the couplings between the QD and the leads to be bilinear and the coupling strength can be described by the hybridization function $\Gamma_{mn}$. Please consult the Methods section for its definition. In the limit of weak couplings and the wide band approximation, we solve the following quantum master equation:\cite{Iv2020,Dou2018} 
\begin{eqnarray}
\partial_t{\hat{\rho}} & = & i[\hat H_{\text{QD}},\hat \rho]-\hat{\hat{\mathcal{L}}}\hat \rho.\label{eq:qme}
\end{eqnarray}
Here, $\hat \rho$ is the density operator for the QD and $\hat{\hat{\mathcal{L}}}$ is a super-operator of Lindblad form that accounts for QD-lead couplings. 

\subsubsection*{Important energy scales}

\begin{figure}[t]
\begin{centering}
\includegraphics[width=14cm]{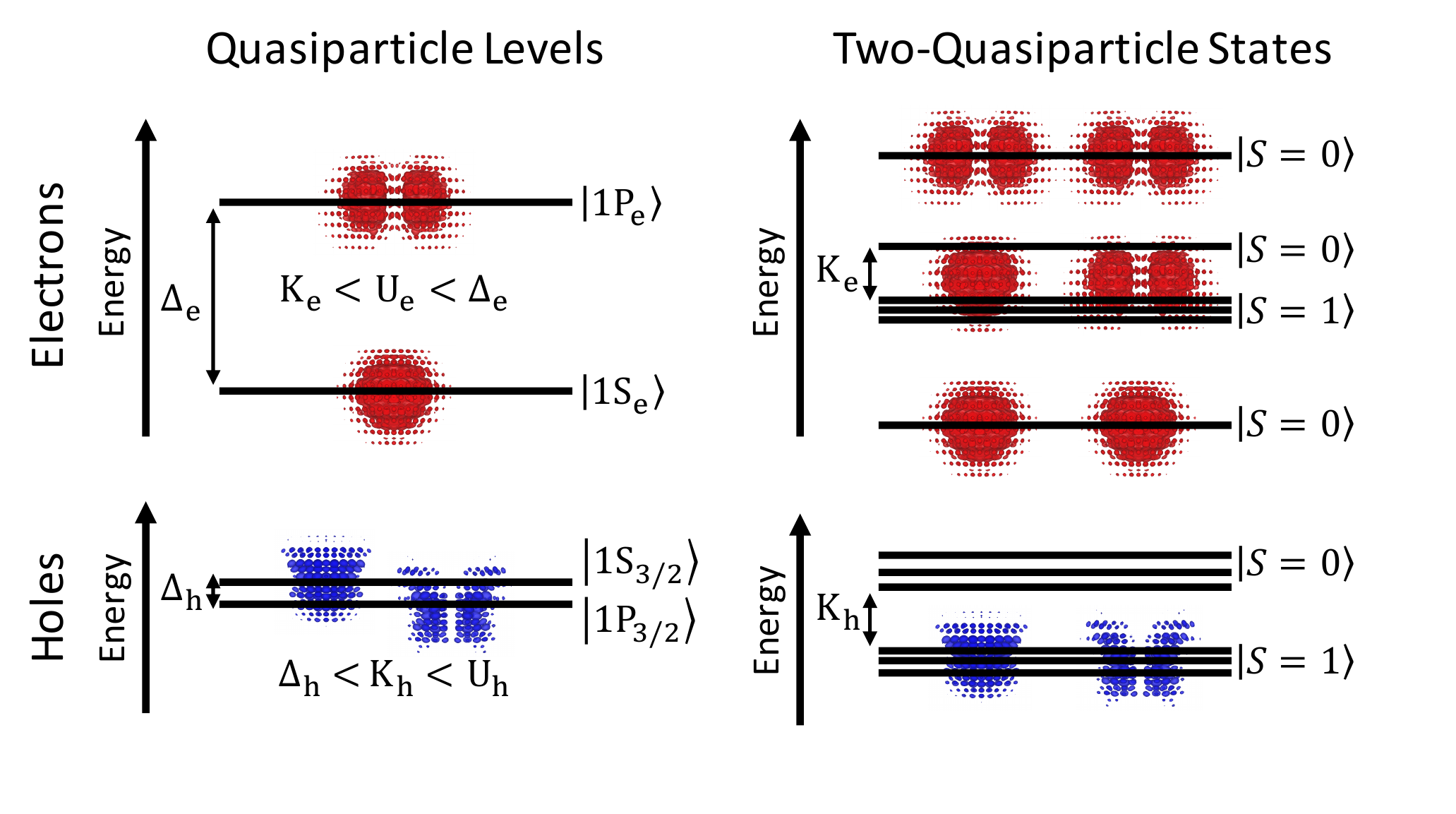}
\par\end{centering}
\centering{}\caption{\label{fig:singlet-triplet-energies}Schematic representation of the energy levels of a quantum dot. (Left column) The the single quasiparticle energy levels (black horizontal lines) are shown for electrons (top) and holes (bottom). The energy spacing between the two lowest lying single quasiparticle energy levels are represented by $\Delta_{\text{e}}$ and $\Delta_{\text{h}}$ for electrons and holes, respectively. Representative charge carrier densities for the electron (red) and hole (blue) states and term symbols are also shown for each single quasiparticle level. The charge carrier densities were taken from a calculation of a CdSe/ZnS core/shell QD with a CdSe core diameter of $2.2$ nm and a ZnS shell of $4$ monolayer thickness. (Right column) The two-quasiparticle states (i.e. two-electron on top and two-hole states on the bottom) are shown along side the spin quantum number ($S$). Neglecting correlations, the lowest lying two-electron state contains both electrons in the $1\text{S}_{\text{e}}$ state forming a spin singlet whereas the lowest lying two-hole state is a spin triplet state formed by placing the holes in different spatial orbitals. This difference in the energetic ordering of spin states in the two-quasiparticle states is a consequence of $K_{\text{e}}<\Delta_{\text{e}}$ and $\Delta_{\text{h}}<K_{\text{h}}$.}
\end{figure}

Before delving into the transport results, it is worthwhile to thoroughly understand the low energy single-particle energy levels ($E_{n}$) and energy scales in the two-quasiparticle states in II-VI QDs. Because of the similar nature of the band structures of III-V and II-VI semiconductors, we speculate that the qualitative results from this discussion should hold for III-V QDs as well. The first energy scale we note is the large energy spacing between the two lowest single electron states ($\Delta_{\text{e}}>100$~meV). The two lowest electron states are termed the 1S\textsubscript{e} and 1P\textsubscript{e} (there are actually three degenerate 1P\textsubscript{e} states for a perfectly spherical QD) states. Representative charge densities for these states are shown in \ref{fig:singlet-triplet-energies} for a CdSe QD. This large energy spacing is a known result, and depends on the QD diameter, ranging from $>500$~meV for CdSe QDs with diameters of $2$~nm to $\sim100$~meV for CdSe QDs with diameters of $10$~nm.\cite{Burda2002} Importantly, the semi-empirical pseudopotential method predicts accurate energy splittings, in quantitative agreement with experimental measurements of this energy splitting.\cite{Jasrasaria2020}

In contrast to electrons, the energy spacing between the two lowest hole states ($\Delta_{\text{h}}$) is very small.\cite{Ekimov1993,Sercel2018} This spacing is on the order of $10$~meV in strongly confined CdSe QDs and approaches the bulk limit of $0$~meV in weakly confined QDs. Thus, this splitting is at least an order of magnitude smaller than the splitting of the electron states. This reduced splitting arises from two primary facts. First, the heavy effective mass of holes in II-VI (and III-VI) semiconductors. Second, the bulk band structure of wurtzite and zincblende II-VI and III-V semiconductors have two (three in the case of negligible spin-orbit coupling) degenerate bands at the valence band maximum. As will be shown next, $\Delta_{\text{h}}$ is the smallest energy scale in the QD Hamiltonian in hole transport (Eq.~\ref{eq:qd-hamiltonian}), which is in stark contrast to $\Delta_{\text{e}}$ being the largest energy scaling in electron transport through QDs (\ref{fig:singlet-triplet-energies}).

The second important energy scale is the Hubbard parameter ($U$). From our calculations, we define $U$ as being equal to $W_{0000}$ where the '$0$' indicates that the spatial orbitals are the lowest unoccupied molecular orbital (LUMO or $1\text{S}_{\text{e}}$) and the highest occupied molecular orbital (HOMO or $1\text{S}_{3/2}$) for electron and hole transport, respectively. For both electrons and holes, the Coulomb repulsion Hubbard parameter ($U\equiv W_{0000}$) is on the order of a couple hundreds of meV for CdSe QDs ranging from $2$~nm to $6$~nm in diameter with or without passivating ZnS shells. This energy scaling is very similar to the energy scale of the exciton binding energy in these systems,\cite{Franceschetti1997,Jasieniak2011,Philbin2018} as expected. 

The third important energy scale in the two-electron and two-hole states is the exchange interaction ($K\equiv W_{0110}$). This Coulomb matrix element $W_{0110}$ has an electron (hole) in the LUMO (HOMO) exchanged with an electron (hole) in the LUMO+1 (HOMO-1). The magnitude of $K$ ranges from $50$~meV to $10$~meV for CdSe QDs ranging from $2$~nm to $6$~nm in diameter. As could have been expected, the magnitude of this splitting is similar to the magnitude of the splitting between the spin singlet and triplet excitons in CdSe QDs.\cite{Sercel2018}

We can summarize these three important energy scales as follows: $K_{\text{e}}<U_{\text{e}}<\Delta_{\text{e}}$ and $\Delta_{\text{h}}<K_{\text{h}}<U_{\text{h}}$ (\ref{fig:singlet-triplet-energies}). In the remaining parts of this work, we will show that these different energetic orderings lead to not only qualitative differences between the nature of the lowest energy two-electron and two-hole states but, also, have important consequences in the spin and charge transport of electrons and holes through QDs, which leads to great implications to the initialization and read-out stages of QD-based quantum computation.

\subsubsection*{Energetic ordering of the two-quasiparticle states}

We now analyze the energy and spin quantum numbers of the two-electron and two-hole states for strongly confined CdSe and CdSe/ZnS QDs. To this end, we employed two computational methods. The first method was to solve the complete QD Hamiltonian (Eq.~\ref{eq:qd-hamiltonian}) within a limited basis of just the two lowest energy single-particle electron and hole states. This method is akin to minimum basis set full configuration interaction calculation, and importantly, reproduces the significant features of our more rigorous approach. Specifically, this method reproduces the level ordering of our second method, that calculates the low energy two-electron and two-hole states by solving an effective Hamiltonian which is very similar to the effective Hamiltonian of an electron-hole pair (i.e. exciton) using the Bethe-Salpeter equation approach.\cite{Rohlfing1998,Rohlfing2000} The major difference between our second method and the Bethe-Salpeter equation is the exchange-like term, which in the former is screened whereas in the latter is not.\cite{Deilmann2016} Please consult the Supporting Information for the explicit expressions of these effective Hamiltonians.

Because of the conceptual simplicity of the first method, we will focus the discussion herein on its predictions, and we refer the interested reader to the Supporting Information for the detailed quantitative predictions of the second method. Upon diagonalization of the QD Hamiltonian (Eq.~\ref{eq:qd-hamiltonian}), using just the two lowest energy single-particle levels (i.e. two electron levels or two holes levels), a qualitative difference is observed between the energetic ordering of the two-electron and two-hole states. In particular, the lowest energy two-electron state is a spin singlet ($\left|S=0\right\rangle $), and the lowest energy two-hole states is a spin triplet ($\left|S=1\right\rangle $). Next, we analyze the complete energetic ordering predicted by this method, with the results depicted in \ref{fig:singlet-triplet-energies}.

The lowest energy two-electron state is, neglecting correlations between the two electrons, simply that of putting both electrons into the LUMO (1S\textsubscript{e}) state of the QD, and in fact, its eigenenergy is reasonably close to $2E_{\text{LUMO}}+U=2E_{0}+W_{0000}$. The energy of this singlet state is significantly lower in energy than the next lowest energy two-electron state due to the single-particle energy gap ($\Delta_{\text{e}}$) being on the order of $500$~meV in strongly confined CdSe and CdSe/ZnS QDs. Specifically, the next two-electron states (again, ignoring correlations) is a set of three degenerate states corresponding to  a spin triplet two-electron configuration ($\left|S=1\right\rangle $), with a single occupancy of both the 1S\textsubscript{e} (LUMO) and 1P\textsubscript{e} (LUMO+1) states. The presence of the exchange interaction ($K_{\text{e}}=W_{0110}$) results in this set of three states being energetically favored relative to the spin singlet state with the same orbital occupations as the set of triplet states, shown schematically in \ref{fig:singlet-triplet-energies}. The last two-electron state in this minimal basis set is that of a spin singlet with double occupation of the 1P\textsubscript{e} state. 

Turning our attention to the low energy two-hole states (bottom right of \ref{fig:singlet-triplet-energies}), we find that the exchange interaction is responsible for the lowest energy two-hole state being a spin triplet state ($\left|S=1\right\rangle $) -- a very important qualitative difference from the lowest energy two-electron state. From an energy scale perspective, this arises for two reasons. First, it is energetically favorable for the holes to occupy different spatial orbitals to reduce the Coulomb repulsion between the two holes, as $W_{0000}>W_{0101}+\Delta_{\text{h}}$. Second, the exchange interaction results in the two-hole configurations with aligned spins being lower in energy than the singlet state with the same spatial orbital occupancy. 

Conceptually, the energetic orderings of the two-electron and two-hole states in strongly confined CdSe QDs (shown in \ref{fig:singlet-triplet-energies}) can be understood and generalized using concepts from introductory science courses such as Hund's rules. For the electron case, the lowest energy quasiparticle levels arise from a single, s-like bulk electronic band with a light effective mass. This results in two-electron configurations being similar to those of the electronic configurations of a helium atom. On the other hand, for the hole case, the lowest energy quasiparticle levels arise from p-like bulk electronic bands with at least one of the bands with a heavy effective mass for II-VI and III-V semiconductor QDs. Thus, the two-hole states are analogous to atomic carbon for which the lowest energy configuration contains two electrons in different p-orbitals and a triplet spin configuration. This connection to atomic orbitals and electronic bands highlights how it is beneficial to utilize principles from both the atomic and bulk material communities to understand the electronic states of semiconductor QDs.

\subsubsection*{Low temperature transport}

Next, we present the calculated low temperature ($k_{B}T\ll U_{\text{e,h}},K_{\text{e,h}},\Delta_{\text{e}}$) charge and spin transport properties of wurtzite CdSe and CdSe/ZnS core/shell QDs, within the strong confinement regime. This size regime together with the low occupancy regime reveals the asymmetric nature of electron and hole transport and highlights the important role of the exchange interaction. We solve Eq.~\ref{eq:qme} using the standard 4\textsuperscript{th} order Runge-Kutta method for both electron and hole transport separately, focusing on the average current ($\left\langle I\right\rangle $) and average spin ($\left\langle S^{2}\right\rangle $) at steady-state as a function of the applied bias, shown schematically in \ref{fig:qd-transport-schematic}. 

\begin{figure}[t]
\centering{}\includegraphics[width=14cm]{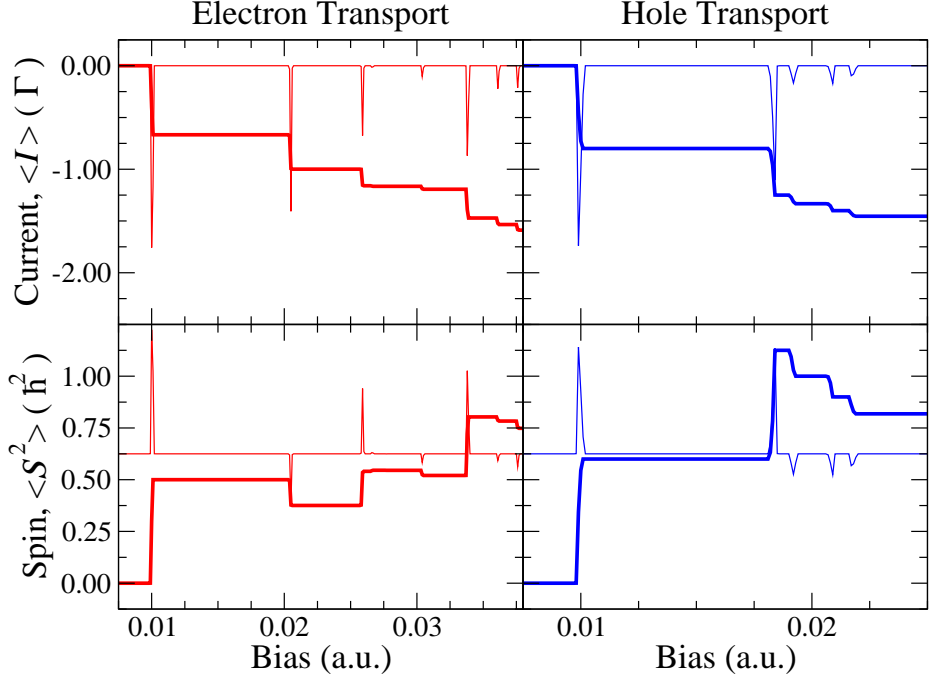}
\caption{\label{fig:transport-with-exchange}Steady-state charge and spin transport spectra for electron (left, red) and hole (right, blue) transport as a function of the voltage bias in atomic units (a.u.), for a CdSe QD with diameter of $3.0$~nm, and leads' temperature of $4$~K. The thick solid lines correspond to the average current ($\left\langle I\right\rangle $) in the top graphs, and spin ($\left\langle S^{2}\right\rangle $) in the bottom graphs. The thin solid lines correspond to the differential current ($d\left\langle I\right\rangle /dV$) and spin ($d\left\langle S^{2}\right\rangle /dV$) in the top and bottom graphs respectively.}
\end{figure}

\ref{fig:transport-with-exchange} presents the average current ($\left\langle I\right\rangle $) and average spin at steady-state ($\left\langle S^{2}\right\rangle $), in thick lines, and their derivatives with respect to the voltage bias ($\frac{d\left\langle I\right\rangle }{dV}$ and $\frac{d\left\langle S^{2}\right\rangle }{dV}$), in thin lines, as a function of the applied bias for a wurtzite CdSe QD with diameter of $3$ nm. Here, the number of single quasiparticle levels is limited to two, as was discussed in the previous section for the two-electron and two-hole states. A step-by-step analysis of all the peaks in \ref{fig:transport-with-exchange} is justified as \ref{fig:transport-with-exchange} constitutes a major finding of this work. 

In our electron (hole) transport simulations, we increase the bias by raising (lowering) the chemical potential of the left lead ($\mu_{L}$) while holding the chemical potential of the right lead ($\mu_{R}$) fixed. Upon ramping up the voltage bias from zero, the steady-state current and spin deviate from zero for the first time once the bias matches the energy of the LUMO and HOMO states for electron and hole transport, respectively. The currents then plateau due to the Coulomb repulsion ($U_{\text{e,h}}$) in both spectra. Additionally, the large energy spacing between the LUMO and LUMO+1 states ($\Delta_{\text{e}}$) compared to the small energy spacing between the HOMO and HOMO-1 states ($\Delta_{\text{h}}$) results in the hole transport plateauing at larger average current and spin values than in the case of electron transport. 

The next peak, located at $0.021$ and $0.018$ atomic units (a.u.) for electron and hole transport, respectively, demonstrate the asymmetry between electron and hole transport and have important consequences in the preparation stage in QD-based quantum computation. Specifically, the bottom panels of \ref{fig:transport-with-exchange} show how the average total spin ($\left\langle S^{2}\right\rangle $) decreases for electron transport, whereas $\left\langle S^{2}\right\rangle $ increases for hole transport within this regime. This asymmetry is a result of the different nature of the lowest energy states in the two-electron and the two-hole state manifolds. The former is a spin singlet, whereas the latter is a spin triplet state (\ref{fig:singlet-triplet-energies}). In terms of state preparation of multi-QD transport\cite{Fujita2017} and QD-based quantum computation, a common initialization scheme is placing two charges on a single QD by increasing the bias until two quasiparticles are on the QD.\cite{Fujita2017,Seedhouse2021} For electrons, this initialization would be done by increasing the bias until $0.023$~a.u. and then quickly removing the bias, such that the two electrons remain on the QD. Because of the large energy spacing between the spin singlet and spin triplet two-electron states, the resulting two electrons on the QD would be in a spin singlet state. On the other hand, if this were done in a hole transport experiment (except the bias was placed at $0.0185$~a.u.), the two holes remaining on the QD would be left in a spin triplet state, since a spin triplet state is the lowest energy two-hole state (as shown schematically in \ref{fig:singlet-triplet-energies}).

For electron transport, the next peaks located at approximately $0.025$~a.u. correspond to the single occupancy of the upper single quasiparticle level (i.e. the 1P\textsubscript{e} level). More interestingly, an exchange induced splitting occurs at a bias of around $0.034$ atomic units. In particular, a large increase in the average spin is shown in the bottom left panel of \ref{fig:transport-with-exchange} at a bias of just over $0.034$. This jump corresponds to the occupancy of the spin triplet state, for which the two electrons occupy different spatial orbitals (\ref{fig:singlet-triplet-energies}). Following the jump, the average spin decreases at bias precisely equal to the exchange energy ($K_{\text{e}}$). The analogue to this in the hole transport is observed by the three peaks around $0.02$ that reduce the average total spin on the QD at steady-state, in accordance with the three singlet states shown in \ref{fig:singlet-triplet-energies}.

\begin{figure}[t]
\centering{}\includegraphics[width=14cm]{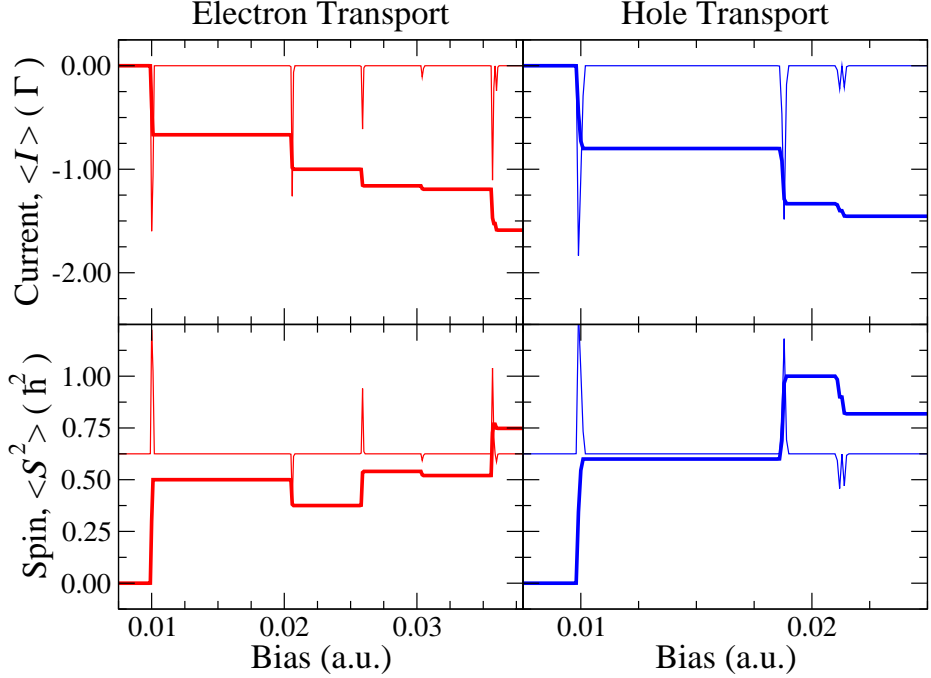}
\caption{\label{fig:transport-no-exchange}Steady-state charge and spin transport spectra for electron (left, red) and hole (right, blue) transport, while neglecting the exchange interactions (i.e. $W_{ijmn}=W_{ijmn}\delta_{im}\delta_{jn}$), as a function of the voltage bias in atomic units (a.u.), for a CdSe QD with diameter of $3.0$~nm, and leads' temperature of $4$~K. The thick solid lines in the top and bottom graphs correspond to the average current ($\left\langle I\right\rangle $) and spin ($\left\langle S^{2}\right\rangle $), respectively. The thin solid lines in the top and bottom graphs correspond to the differential current ($d\left\langle I\right\rangle /dV$) and spin ($d\left\langle S^{2}\right\rangle /dV$), respectively.}
\end{figure}

The exact role of the exchange interaction in \ref{fig:transport-with-exchange} can be clearly seen by analyzing spectra in which we ``turn off'' the exchange interaction. Mathematically, this is achieved by setting $W_{ijmn}=W_{ijmn}\delta_{im}\delta_{jn}$. \ref{fig:transport-no-exchange} shows the calculated spectra for which exchange effects were neglected. The first thing in \ref{fig:transport-no-exchange} to note is that there are less peaks relative to \ref{fig:transport-with-exchange}. In this case, the exchange interaction does not lift the degeneracy between spin singlet and spin triplet states with the same spatial occupancy.

Specifically, two of the peaks that occur around $0.035$ and $0.018$~a.u. in the electron and hole transport spectra, respectively, merge when exchange interactions are neglected. Furthermore, the lifting of the degeneracy between spin singlets and triplets at these biases by the exchange interaction in \ref{fig:transport-with-exchange} results in first a large increase in the average spin due to the occupation of the spin triplet states. This increase in the average spin is quickly followed by a decrease in the average spin due to the occupation of the singlet state at a slightly larger bias. In \ref{fig:transport-no-exchange}, the lack of the exchange interaction results in these two peaks combining with the net effect of a small increase in the average spin. Thus, we have shown how spin-exchange blockades can increase the number of peaks in a spectra and how their locations in the spectra depend on both the magnitude of the exchange interaction ($K_{\text{e,h}}$) and the density of states near the band edge ($\Delta_{\text{e,h}}$).

\subsubsection*{Temperature and size dependent transport}

Because we build the QD Hamiltonian in Eq.~\ref{eq:qd-hamiltonian} from atomistic electronic structure calculations of realistically sized CdSe and CdSe/ZnS QDs, we are not only able to elucidate new physics from the use of accurate single quasiparticle energies ($E_{n}$) and Coulomb matrix elements ($W_{ijmn}$), but we can also study the temperature and size dependence of these results. We begin this subsection by discussing the temperature dependence of the features discussed in detail in the previous section, for a CdSe QD with a diameter of $3$~nm. The results presented here (and above) are valid also for the case of CdSe QDs with insulating ZnS shells, as demonstrated in the Supporting Information. This point is of great practical importance, as shell thickness and composition are valuable knobs that can be controlled to optimize many properties of the overall nanomaterial.\cite{Klimov2007,Chen2013,Hadar2017,Hanifi2019} For example, in terms of transport studies, increasing the shell thickness should decrease the metal-QD coupling (i.e. system-bath coupling) while hardly altering the lowest energy single-particle electron and hole states, because they remain entirely within the CdSe core in CdSe/ZnS core/shell QDs.\cite{Philbin2020}

\begin{figure}[t]
\centering{}\includegraphics[width=14cm]{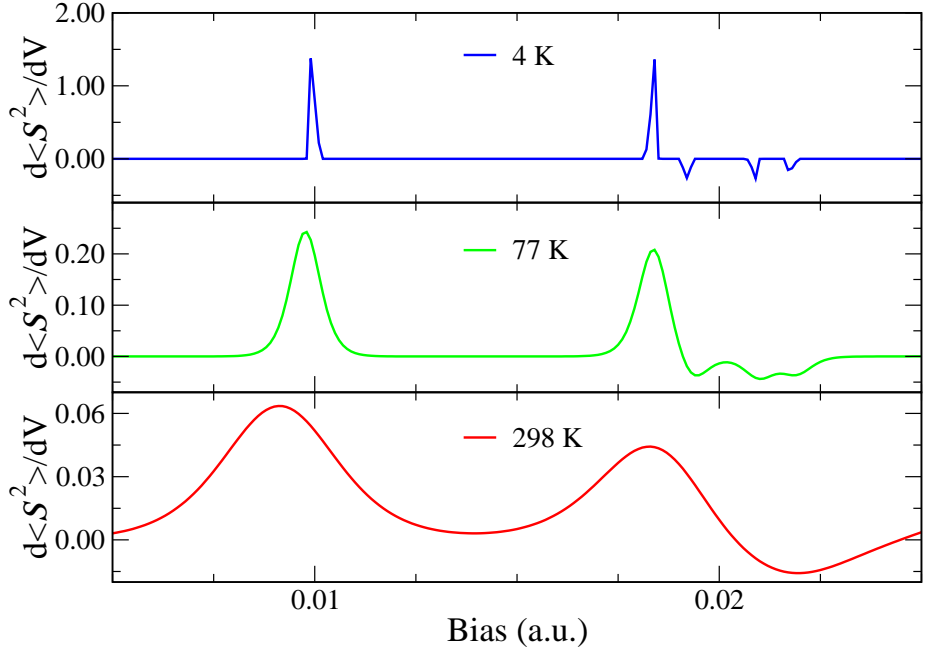}
\caption{\label{fig:temperature-dependence}Temperature dependence of the derivative of the average spin at steady-state ($d\left\langle S^{2}\right\rangle /dV$) as a function of voltage bias, for hole transport. The full QD Hamiltonian was used for a CdSe QD with diameter of $3.0$~nm, and leads' temperature of $4$~K, $77$~K, and $298$~K from top to bottom.}
\end{figure}

\ref{fig:temperature-dependence} shows that at $4$~K, the peaks in the $d\left\langle S^{2}\right\rangle /dV$ curves which are determined by the density of states, Coulomb repulsion, and exchange interaction, are very sharp. Specifically, the hole transport shows a large second peak that increases the total spin, followed by three smaller negative peaks that arise from the three spin singlet two-hole states. These peaks are slightly higher in energy than the spin triplet, as discussed above, and shown in \ref{fig:singlet-triplet-energies}. Upon increasing the temperature to $77$~K, the peaks begin to smear out, but yet the overall features still remain. Further increasing the temperature to $298$~K, these peaks have begun merging. 

To study the size dependence, we present similar spectra as shown above for CdSe QDs ranging from $d=2$ to $d=4$ with up to $4$ monolayers of ZnS shell material in the Supporting Information. Importantly, all of the qualitative conclusions drawn from the specific spectra shown above do not change upon increasing or decreasing the size of the QD. There are a few quantitative changes worth noting though. First, the magnitude of the voltage bias needed for adding two charges to the system decreases with increasing system size. Similarly, the magnitude of the spacing between the peaks decreases upon increasing QD size. Thus, observing all the peaks for larger QDs, which implies smaller energy scales of $\Delta_{\text{e,h}}$, $U_{\text{e,h}},$ and $K_{\text{e,h}}$, requires lower temperatures. That being said, all of the peaks remain sharp up to $4$~K, even for the largest QD studied herein. This is a promising result for the future of ``high temperature'' QD-based quantum computation and transport studies.

\subsection*{Conclusion}

In this work, we systematically studied charge and spin transport through strongly confined QDs by parameterizing a quantum master equation from atomistic electronic structure calculations. This unique combination of electronic structure calculations and open quantum system methods allowed us to build accurate QD Hamiltonians for experimentally relevant QDs containing hundreds of atoms. By modelling the QDs connected to two leads with a voltage bias, we found that electron and hole transport exhibit very different spin-dependent transport properties. We explained their differences and locations of the spin-dependent blockades by detailing the low energy two-electron and two-hole states. In particular, we showed that as a result of the interplay of the single quasiparticle density of states, Coulomb repulsion energies, and exchange interactions, the lowest energy two-electron state is a spin singlet, whereas the lowest energy two-hole state is a spin triplet.

Because spin-dependent blockades have important implications for the initialization and read-out stages of quantum transport and computation applications, the methodology introduced in this work provides a systematic way to combine atomistic electronic structure calculations with model Hamiltonian methods to understand and, eventually, design future experiments. A few particularly exciting future directions include the extension of this work to multiple QD systems,\cite{Bayer2001,Mills2019,Cui2019,Panfil2019,Chan2021} how the spin states are impacted by phonons,\cite{Kershaw2018,Erpenbeck2020} and how the spin states can be manipulated using photons.\cite{Imamoglu1999}

\subsection*{Methods}

Instead of using a model Hamiltonian (e.g. Eq.~\ref{eq:qd-model-hamiltonian}) to represent the CdSe and CdSe/ZnS core/shell QDs, we calculate the electron and hole states using atomistic electronic structure methods for experimentally relevant colloidal QDs. Due to the large sizes of QDs ($>1,000$ electrons), we cannot readily use fully \textit{ab-initio} methods (e.g. density functional theory or many-body perturbation theory) to obtain these quasiparticle states due to their high computational cost. We circumvent this high computational cost by utilizing the semi-empirical pseudopotential method\cite{Wang1996,Rabani1999b} along with filter-diagonalization techniques.\cite{Wall1995,Toledo2002} In doing so, we calculate the low energy electron and hole states with reduced computational cost, which enables us to explore the transport properties of QDs with more than $1,000$ valence electrons. Within the semi-empirical pseudopotential method, we solve the following time-independent Schr\"{o}dinger equation:
\begin{eqnarray}
\left[-\frac{1}{2}\nabla^{2}+V_{ps}\left(\mathbf{r}\right)\right]\psi_{n}\left(\mathbf{r}\right) & = & E_{n}\psi_{n}\left(\mathbf{r}\right).\label{eq:methods-sepm-hamiltonian}
\end{eqnarray}
In Eq.~\ref{eq:methods-sepm-hamiltonian}, the two first terms on the left hand side are the kinetic energy ($-\frac{1}{2}\nabla^{2}$) and the pseudopotential potential energy ($V_{ps}\left(\mathbf{r}\right)$), and $\psi_{n}$ and $E_{n}$ are the quasiparticle wavefunction and energy, respectively. Filter-diagonalization techniques\cite{Wall1995,Toledo2002} were utilized to efficiently solve for just the low energy electron and hole states that are central to the lowest energy spin singlet and triplet two-electron and two-hole states in QDs. This semi-empirical pseudopotential approach has been successfully applied to understand the electron, hole, and electron-hole pair (i.e. excitonic) states in both III-V and II-VI semiconductor QDs very well.\cite{Wang1996,Rabani1999b,Williamson2000,Philbin2018,Philbin2020}

To account for the many-body interactions between the quasiparticles, we calculate screened Coulomb matrix elements, 
\begin{eqnarray}
W_{ijmn} & = & \int\int\,\psi_{i}^{*}\left(x\right)\psi_{j}^{*}\left(y\right)W\left(x,y\right)\psi_{m}\left(x\right)\psi_{n}\left(y\right)\,dx\,dy\,,\label{eq:methods-coulomb-matrix-elements}
\end{eqnarray}
using the atomistic wavefunctions ($\psi_{n}$) obtained from solving Eq.~\ref{eq:methods-sepm-hamiltonian} and approximating the screened Coulomb potential ($W\left(x,y\right)$) using the static dielectric approximation with a dielectric constant $\epsilon=5$. The Coulomb matrix elements given by Eq.~\ref{eq:methods-coulomb-matrix-elements} contain both the direct Coulomb repulsion and exchange effects, providing more accurate and realistic interaction strengths between the quasiparticles compared to having a single Hubbard parameter ($U$) and exchange parameter ($K_{S}$) as typically used in QD model Hamiltonians such as Eq.~\ref{eq:qd-model-hamiltonian}.

Having determined the quasiparticle energies ($E_{n}$) and the screened Coulomb matrix ($W_{ijmn}$), we can now build a Hamiltonian for the QDs ($\hat H_{\text{QD}}$) as 
\begin{eqnarray}
\hat{H}_{\text{QD}} & = & \sum_{n}E_{n}\hat{d}_{n}^{\dagger}\hat{d}_{n}+\sum_{ijmn}W_{ijmn}\hat{d}_{i}^{\dagger}\hat{d}_{j}\hat{d}_{n}^{\dagger}\hat{d}_{m}\label{eq:methods-qd-hamiltonian}
\end{eqnarray}
where $\hat{d}_{n}$ ($\hat{d}_{n}^{\dagger}$) is the annihilation (creation) operator for quasiparticle state $\psi_{n}$ with energy $E_{n}$. Now that we have an accurate Hamiltonian for the QDs based on atomistic electronic structure theory, we can couple the QD to two leads in order to study the transport properties through the QDs (as shown schematically in \ref{fig:qd-transport-schematic}).

The total Hamiltonian ($\hat H$) consists of three parts: $\hat H_{\text{M}}$ representing the two metal leads, $\hat H_{\text{QD}}$ representing the semiconductor QD, and $\hat H_{\text{I}}$ containing all the coupling between the metal leads and the QD, i.e. 
\begin{eqnarray}
\hat H & = & \hat H_{\text{M}}+ \hat H_{\text{QD}}+ \hat H_{\text{I}}.
\end{eqnarray}
Each metal lead Hamiltonian $\hat{H}_{\text{M}}=\sum_{k\alpha}\epsilon_{k\alpha}\hat{c}_{k\alpha}^{\dagger}\hat{c}_{k\alpha}$, consists of a manifold of electronic states indexed by $k$, and the QD-leads interaction Hamiltonian $\hat{H}_{\text{I}}=\sum_{nk}V_{nk}(\hat{c}_{k}^{\dagger}\hat{d}_{n}+\hat{d}_{n}^{\dagger}\hat{c}_{k})$ is assumed to be bilinear. To quantify the interaction strength, we define the hybridization function: 
\begin{eqnarray}
\Gamma_{mn}(\epsilon) & = & \sum_{k}V_{mk}V_{nk}\delta(\epsilon_{mn}-\epsilon).
\end{eqnarray}
In this work, we assume the wide-band approximation such that the hybridization function is independent of $\epsilon$. We will further assume that the QD-lead couplings are weak (i.e. $\Gamma_{mn}<k_{B}T$ where $k_{B}$ is the Boltzmann constant and $T$ is the temperature of the entire system). This assumption allows us to trace over the electronic states in the metal leads and use a quantum master equation\cite{Nitzan2006,Breuer2002} to analyze the reduced dynamics of the QD, 
\begin{eqnarray}
\partial_t{\hat{\rho}} & = & i[\hat H_{\text{QD}},\hat \rho]-\hat{\hat{\mathcal{L}}}\hat \rho,\label{eq:methods-qme}
\end{eqnarray}
where $\hat \rho$ is the many-body density operator for the system and $\hat{\hat{\mathcal{L}}}$ is a super-operator of Lindblad form that accounts for the system-bath interactions.\cite{Lindblad1976,Gorini1976} Please consult previous works for more details on the methodology.\cite{Dou2016,Dou2017}

\begin{acknowledgement}
This work is primarily supported by the Department of Energy, Photonics at Thermodynamic Limits Energy Frontier Research Center under Grant No. DE-SC0019140. J.P.P. is a Ziff Fellow at the Harvard University Center for the Environment. P. N. is a Moore Inventor Fellow through Grant GBMF8048 from the Gordon and Betty Moore Foundation. This research used resources of the National Energy Research Scientific Computing Center (NERSC), a U.S. Department of Energy Office of Science User Facility operated under Contract No. DE-AC02-05CH11231.
\end{acknowledgement}

\bibliography{quantumDotTransport}

\end{document}